\documentclass[conference]{IEEEtran}
\IEEEoverridecommandlockouts
\usepackage{amsmath,amssymb,amsfonts}
\usepackage{algorithmic}
\usepackage{textcomp}
\usepackage{bm}
\usepackage{graphicx}
\usepackage{booktabs}
\usepackage{xcolor}
\usepackage{hyperref}
\usepackage{subcaption}
\def\BibTeX{{\rm B\kern-.05em{\sc i\kern-.025em b}\kern-.08em
    T\kern-.1667em\lower.7ex\hbox{E}\kern-.125emX}}

\usepackage[backend=biber,style=ieee]{biblatex}
\AtEveryBibitem{\clearfield{language,url,eprint}}
\AtEveryBibitem{\clearfield{url}}
\AtEveryBibitem{\clearfield{eprint}}

\AtEveryBibitem{\clearlist{language}} 
\AtEveryBibitem{\clearlist{url}} 
\AtEveryBibitem{\clearlist{eprint}} 

\usepackage{tikz}
\newcommand\copyrighttext{%
	\footnotesize This work has been submitted to the IEEE for possible publication. 
	Copyright may be transferred without notice, after which this version may no longer be accessible.}
\newcommand\copyrightnotice{%
	\begin{tikzpicture}[remember picture,overlay]
		\node[anchor=south,yshift=10pt] at (current page.south) {\fbox{\parbox{\dimexpr\textwidth-\fboxsep-\fboxrule\relax}{\copyrighttext}}};
	\end{tikzpicture}%
}

\begin{document}

\title{Embedded DNA Inference in In-Body Nanonetworks: Detection, Delay, and Communication Trade-Offs}



\author{%
\IEEEauthorblockN{Stefan Fischer}
\IEEEauthorblockA{%
Institute for Telematics, University of L\"ubeck, Germany \\
Email: stefan.fischer@uni-luebeck.de
}
}


\maketitle
\copyrightnotice 

\begin{abstract}
In-body molecular nanonetworks promise early abnormality detection close to the source of biochemical events, but their communication capabilities are severely constrained by slow diffusion-based signaling and unstable alarm traffic. We study whether simple embedded DNA-based inference at the nanonode can improve alarm transmission to an external gateway. We compare raw reporting (RR), single-marker threshold reporting (TR), and embedded inference reporting (EIR) under a communication-oriented abstraction of DNA strand-displacement-based computation with marker gating, edge-triggered alarming, hysteretic state transitions, temporally correlated marker dynamics, diffusion-based alarm transport, and leaky gateway evidence integration. The simulations identify a bounded EIR success regime in the weak-to-moderate anomaly range: EIR can improve detection relative to RR and TR while remaining competitive in event-driven communication cost, especially relative to RR. The gain does not come from uniformly lower activity, but from more stable local alarm dynamics. EIR does not dominate globally; TR often remains cheaper when abnormalities are present, and EIR incurs additional local delay. These results point to a limited operating regime in which EIR is useful, rather than to a general advantage across settings.
\end{abstract}


\section{Introduction}

Recent years have seen growing interest in \emph{Physical Neural Networks} (PNNs), i.e., systems that realize neural computation directly in physical substrates rather than solely through conventional digital hardware. As discussed in our recent survey, this trend is driven by increasing energy and data-movement constraints in silicon-centered AI, especially in embedded and edge settings \cite{Fischer2026PNNSurvey}. Within this broader landscape, DNA-based molecular neural systems are particularly relevant when sensing and computation must operate directly on biochemical signals.

Against this background, molecular nanonetworks are a promising paradigm for in-body sensing and early abnormality detection, as they can operate close to the biochemical source of an event and communicate through molecules in environments where conventional electromagnetic communication is ineffective or impractical \cite{torresgomez2024abnormality}. However, molecular communication is slow, bandwidth-limited \cite{kuran2021}, and easily burdened by excessive alarm traffic when many nanonodes monitor weak and noisy biochemical signatures \cite{gao2024molecularcodedivisionmultipleaccesssignaling}. A straightforward reporting strategy, in which each nanonode forwards every locally relevant observation or every single-marker threshold crossing, is therefore often inefficient. It can overload the gateway with redundant or weakly informative alarms, increase false positives, and waste scarce communication opportunities in a strongly constrained in-body environment. The key difficulty is therefore not sensing alone, but deciding which local observations are important enough to justify molecular transmission.

While prior work has considered abnormality detection with local sensing and fusion-center or gateway-based decision making in molecular nanonetworks, the explicit use of an embedded molecular inference block at the nanonode to suppress redundant alarms, and the resulting trade-off between reduced communication load and additional local inference delay, appears to have received little direct attention.

DNA-based molecular computing offers one possible way to address this problem directly at the nanonode. In particular, DNA strand displacement (DSD) enables programmable biochemical computation in which signal strands trigger the release or suppression of downstream strands \cite{Simmel2023-eq}. Prior work has shown that such systems can implement concentration-based signal processing, weighted summation, and nonlinear decision functions via sequestration and winner-take-all mechanisms \cite{CherryQian2018,QianWinfree2014,cherry2025supervised}. This makes DNA-based local inference a plausible candidate for communication-aware pre-processing of multi-marker observations before alarm emission.

However, local DNA inference is not free. Although DNA substrates offer extreme parallelism and direct processing of molecular inputs, their effective operating bandwidth remains limited, and complex strand-displacement cascades may require minutes or even hours to approach equilibrium \cite{CherryQian2018}. The central question is therefore under which operating conditions reduced alarm traffic and false alarms outweigh the additional local reaction delay.

In this paper, we investigate communication-aware embedded DNA inference for in-body molecular nanonetworks. We consider nanonodes that sense multiple biochemical markers and perform a local molecular decision step before transmitting an alarm to an external gateway. Rather than claiming a sequence-level biochemical realization, we adopt a communication-oriented functional abstraction consistent with DNA-based concentration weighting, thresholding, and competitive decision mechanisms. Based on this abstraction, we compare raw reporting (RR), single-marker threshold reporting (TR), and embedded inference reporting (EIR).

The contributions of this paper are threefold: (i) we propose a compact system model for in-body molecular nanonetworks with embedded DNA-inspired local inference, diffusion-based alarm transport, and leaky gateway evidence integration; (ii) we compare RR, TR, and EIR under a reporting model that explicitly captures marker gating, edge-triggered alarming, hysteretic local state dynamics, variance-preserving temporally correlated marker evolution, and operating-point calibration under absence of abnormalities; and (iii) we show by simulation that EIR admits a bounded success regime in which detection improves in the weak-to-moderate anomaly range while communication cost remains competitive rather than uniformly minimal.

\section{Related Work}

\subsection{Molecular Communication and In-Body Nanonetworks}
Molecular communication and in-body nanonetworking have been studied as enabling paradigms for communication among nanoscale devices in environments where conventional electromagnetic signaling is ineffective or impractical. Prior survey and vision papers established the general communication-theoretic foundations of diffusion-based molecular channels, highlighted their relevance for nanomedicine, and framed the broader Internet of Bio-Nano Things (IoBNT) perspective \cite{Farsad2016,Atakan2012,Akyildiz2015}. This body of work has clarified both the promise and the main limitations of the paradigm, including slow propagation, channel memory, inter-symbol interference, and the difficulty of scaling communication reliably when many nanonodes share the same biochemical environment.

A closely related line of work has addressed abnormality detection in molecular nanonetworks using distributed sensing nodes and a sink-side decision entity. Early two-tier schemes considered sensor nano-machines that locally detect abnormalities and report to a data-gathering node for final decision making \cite{Ghavami2012,Ghavami2017}. Subsequent work studied cooperative abnormality detection with several sensors and a fusion center, including the analysis of reporting schemes, fusion rules, and detection performance under diffusive channels \cite{Mosayebi2017,Varshney2018}. More recent work has also considered gateway-related system aspects such as alarm-system architectures and timeliness-aware abnormality detection in the human circulatory system \cite{lau2025using,pal2024age}.

However, these works primarily focus on sensing, reporting, fusion, gateway placement, or detection timeliness rather than on the question central to this paper: whether an embedded molecular inference block at the nanonode can suppress weak or redundant alarms before transmission. In particular, the explicit trade-off between reduced communication load and the additional delay introduced by local molecular inference appears to have received little direct attention in the in-body nanonetworking literature.

\subsection{DNA Computing and Molecular Neural Systems}
DNA strand displacement (DSD) has emerged as a central mechanism for programmable molecular computation and provides the biochemical basis for implementing signal processing directly in the molecular domain. Reviews of DSD-based computational systems summarize how logic gates, cascades, and larger reaction networks can be constructed from sequence-programmable strand interactions \cite{Simmel2023-eq,Chen2023}. A key milestone was the demonstration that DSD cascades can realize neural-network-style computation, including weighted summation and thresholding, in molecular form \cite{QianWinfree2011}. Closely related work also showed that scalable digital circuit computation can be implemented using strand displacement cascades \cite{QianWinfree2014}.

Subsequent research considerably expanded the expressive power of molecular neural systems. Winner-take-all DNA neural networks enabled large-scale molecular pattern recognition \cite{CherryQian2018}, while more recent work demonstrated deeper fully connected and convolutional DNA neural networks \cite{Liu2025} as well as supervised learning in DNA neural networks \cite{cherry2025supervised}. Beyond non-enzymatic DSD systems, enzymatic molecular neural networks have also shown strong nonlinear decision-making capabilities \cite{Okumura2022-jx}. Nevertheless, this line of work is primarily concerned with realizing molecular computation itself. It does not directly address how such molecular inference blocks should be integrated into communication-constrained in-body nanonetworks in order to reduce unnecessary transmissions.

\subsection{Local Multi-Marker Decision Making}
A closely related literature studies molecular logic and biochemical decision systems for multi-input sensing and classification. DNA-based molecular computation has been used for multi-input diagnostic classification and the molecular analysis of complex biomarker profiles \cite{Lopez2018,Zhang2020}. More recently, spatially localized DNA classifiers have shown that localized molecular decision systems can execute programmable classification of multiple miRNA inputs in a faster and more effective manner in clinical samples \cite{Yang2024}. In addition, multi-threshold and multi-input DNA logic designs have been proposed for profiling cancer-related microRNA signatures \cite{Sanjabi2019}, and recent review work has highlighted the growing practical relevance of DNA computing for diagnostic applications and molecular readout technologies \cite{Takiguchi2025}.

These studies show that combining multiple molecular inputs can improve selectivity and enable richer decision rules than single-marker thresholding alone. However, their primary objective is typically diagnostic classification accuracy or biochemical functionality rather than communication efficiency at the nanonetwork level. In contrast, our focus is on local multi-marker decision making as a communication-aware mechanism: the goal is not only to improve selectivity, but to suppress weak or redundant alarms before molecular transmission. Consistent with recent survey literature showing that neural-network-related work in molecular communication is still dominated by PHY-layer tasks such as channel estimation, synchronization, and detection, the coupled trade-off between local molecular inference delay, false-alarm suppression, and alarm-traffic reduction appears to have received little direct attention in the in-body nanonetworking literature \cite{Gomez2025CommunicatingSmartly}.

\section{System Model}

\subsection{Considered Scenario}

We consider an in-body molecular nanonetwork deployed in a bounded fluidic environment such as a vessel segment or a localized interstitial region. A set of $N$ nanonodes is distributed in this environment and continuously monitors the local biochemical state. An abnormality event, such as the local emergence of a pathological process, changes the concentrations of one or more biochemical markers. The objective of the nanonetwork is to detect such an event and report it to an external gateway by means of molecular signaling.

Each nanonode senses a vector of $m$ biochemical marker concentrations,
\begin{equation}
\mathbf{x}_i(t) = [x_{i,1}(t), x_{i,2}(t), \ldots, x_{i,m}(t)],
\end{equation}
where $x_{i,j}(t)$ denotes the local concentration of marker $j$ at node $i$ and time $t$. In the present paper, we focus on the practically relevant case $m=2$, which already captures the central trade-off between a sensitive but unspecific marker and a more specific but weaker marker.

The gateway is placed outside the immediate nanonetwork domain or at its boundary and receives alarm molecules emitted by nanonodes. Based on the arriving alarm pattern, the gateway decides whether an abnormality event is present.

\subsection{Nanonode Architecture}

Each nanonode consists of four functional blocks:
\begin{enumerate}
    \item \textbf{Sensing block:} The sensing block measures the local biochemical marker concentrations.
    \item \textbf{Local inference block:} The local inference block combines the sensed marker concentrations before transmission. We do not model this block at sequence level. Instead, we use a functional abstraction consistent with DNA strand-displacement-based molecular computation, in which marker concentrations act as analog inputs, intermediate gate concentrations realize weighting, and sequestration or competitive reactions implement nonlinear decision making.
    \item \textbf{Alarm release block:} If the local decision exceeds a predefined threshold, the node releases alarm molecules into the environment.
    \item \textbf{Communication block:} The released alarm molecules propagate toward the gateway through diffusion, optionally superimposed with drift or flow, depending on the considered environment.
\end{enumerate}

\subsection{Local Inference Model}

For each node $i$, the local inference block computes a scalar decision variable
\begin{equation}
z_i(t)=\sum_{j=1}^{m} w_j x_{i,j}(t)-\theta,
\end{equation}
where $w_j$ denotes the effective contribution of marker $j$ and $\theta$ is the local decision threshold. In the present instantiation with $m=2$, EIR combines a sensitive but less specific marker with a weaker but more specific second marker. A positive EIR state requires both a sufficiently large score and a sufficiently large concentration of the second marker. With hysteresis, ON and OFF transitions are separated:
\begin{align}
\text{ON:}\quad & z_i(t)>\theta,\; x_{i,2}(t)>\tau_g,\\
\text{OFF:}\quad & z_i(t)<\theta-\Delta_z\; \text{or}\; x_{i,2}(t)<\tau_g-\Delta_g,
\end{align}
where $\tau_g$ is the marker-2 gate threshold, $\Delta_z$ the score hysteresis margin, and $\Delta_g$ the gate hysteresis margin. This realizes a specificity gate together with stabilized local state transitions.

This abstraction is motivated by three experimentally and conceptually established properties of DNA-based molecular neural systems: (i) concentration-based signal encoding, (ii) weighting via reaction stoichiometry or gate concentrations, and (iii) nonlinear thresholding via sequestration or competitive hybridization. The abstraction allows us to analyze communication-system behavior without claiming a full biochemical realization of the underlying DNA circuit\footnote{There is a number of potential existing DNA technologies to be used for the real implemenantion. We note however, that this is not the purpose of this paper which instead aims at establishing more general results. Concrete realizations are postponed to future work.}.

\subsection{Reporting Strategies}

We compare three reporting strategies.

\textit{Raw reporting (RR):} Each nanonode emits an alarm whenever a locally sensed marker becomes relevant according to a basic sensing rule. This strategy minimizes local processing but may generate many transmissions and false alarms.

\textit{Single-marker threshold reporting (TR):} Each nanonode emits an alarm only if a single designated marker exceeds a predefined threshold. This reduces some communication overhead but cannot exploit multi-marker evidence.

\textit{Embedded inference reporting (EIR):} Each nanonode first combines multiple marker concentrations using the local inference block and emits an alarm only if the inferred decision output becomes positive. This strategy introduces additional local reaction delay but may substantially reduce alarm traffic and improve specificity.

\subsection{Timing, Communication, and Event Model}

The end-to-end detection process at node $i$ consists of three delay components:
\begin{itemize}
    \item sensing delay $T_s$,
    \item local inference delay $T_i$,
    \item communication delay $T_c$.
\end{itemize}
For raw reporting, the node does not perform local inference, and the reporting delay is
\begin{align}
D_i^{\mathrm{RR}} &= T_s + T_c, \\
D_i^{\mathrm{TR}} &\approx T_s + T_c, \\
D_i^{\mathrm{EIR}} &= T_s + T_i + T_c.
\label{eq:eir_delay}
\end{align}
The additional term $T_i$ is central to this study because communication savings are only meaningful if they are not outweighed by excessive local reaction time.

When a node emits an alarm, it releases $M_a$ alarm molecules into the environment. For node $i$ at distance $d_i$ from the gateway, the contribution of one alarm emission is modeled by a one-dimensional diffusion-based impulse response
\begin{equation}
h_i(t)=\eta\,\frac{1}{\sqrt{4\pi D(t+\varepsilon)}}\exp\!\left(-\frac{(d_i-vt)^2}{4D(t+\varepsilon)}\right)e^{-\lambda t},\qquad t>0,
\end{equation}
where $D$ denotes the diffusion coefficient, $v$ an optional drift velocity, $\lambda$ an optional decay rate, and $\varepsilon$ a small regularization term induced by the finite receiver size. In the present evaluation we use pure diffusion ($v=0$) without molecular decay ($\lambda=0$). The gateway does not decide on instantaneous arrivals alone. Instead, it maintains a leaky evidence state
\begin{equation}
A_k = \beta A_{k-1} + E_k,\qquad \beta=e^{-\Delta t/\tau_g^{(\mathrm{gw})}},
\end{equation}
where $E_k$ is the alarm evidence arriving in time step $k$ and $\tau_g^{(\mathrm{gw})}$ is the gateway integration time constant. A gateway alarm is declared when $A_k$ exceeds a predefined detection threshold $\Theta_g$.

We distinguish between the environmental states $\mathcal{H}_0$ (no abnormality) and $\mathcal{H}_1$ (abnormality present). Under $\mathcal{H}_1$, the local marker vector at a subset of nodes changes according to an anomaly signature. In the simplest case, marker~1 exhibits a strong but less specific increase, whereas marker~2 exhibits a weaker but more specific increase. Under $\mathcal{H}_0$, marker concentrations fluctuate due to physiological background activity and sensing noise, which may still trigger false alarms under insufficiently selective reporting rules. To capture biochemical persistence without distorting the nominal per-marker variance, we generate temporally correlated marker traces as a mean-tracking AR(1)-type process,
\begin{equation}
\begin{split}
\tilde{x}_{i,j}(t_k) = {} & \mu_{i,j}(t_k)
+ \alpha \big(\tilde{x}_{i,j}(t_{k-1}) - \mu_{i,j}(t_{k-1})\big) \\
& {} + \sqrt{1-\alpha^2}\,\sigma_j\,\xi_{i,j}(k),
\end{split}
\end{equation}
with $\xi_{i,j}(k)\sim\mathcal{N}(0,1)$ and $x_{i,j}(t_k)=\max\{0,\tilde{x}_{i,j}(t_k)\}$. This yields temporally persistent but variance-controlled biochemical evidence.

\begin{figure}[htbp]
  \centering
  \includegraphics[width=\linewidth]{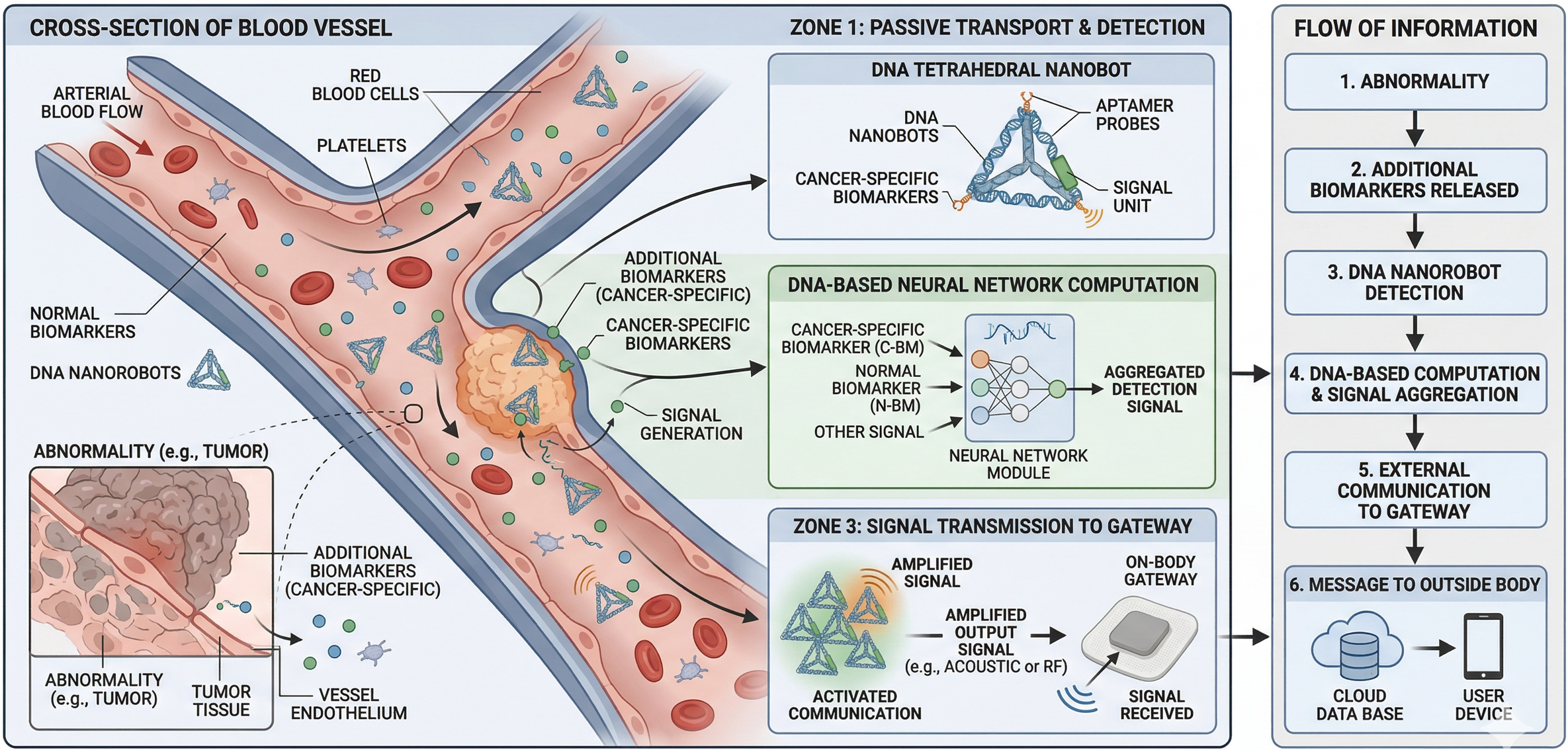} 
\caption{Conceptual framework for in-vivo abnormality detection and external communication via DNA nanorobots (Section III). Left: Blood vessel cross-section illustrating DNA nanobots detecting cancer-specific biomarkers. Center: Functional stages including passive detection, DNA-based neural network inference for signal aggregation, and signal transmission to an on-body gateway. Right: Sequential flow of information from initial biochemical detection to external data reporting on a user device.}
  \label{fig:system_model}
\end{figure}

The overall system model is visualized in Fig.~\ref{fig:system_model}.

\section{Performance Metrics and Design Hypothesis}

For strategy $s\in\{\mathrm{RR},\mathrm{TR},\mathrm{EIR}\}$, we use five primary metrics:
\begin{align}
P_D^{(s)} &= \Pr\big(\widehat{\mathcal{H}}=\mathcal{H}_1\mid \mathcal{H}_1,s\big), \\
P_{FA}^{(s)} &= \Pr\big(\widehat{\mathcal{H}}=\mathcal{H}_1\mid \mathcal{H}_0,s\big), \\
C_{H_0}^{(s)} &= M_a\sum_{i=1}^{N} n_i^{(s)}([0,T]\mid \mathcal{H}_0), \\
C_{H_1}^{(s)} &= M_a\sum_{i=1}^{N} n_i^{(s)}([t_0,t_{\mathrm{det}}]\mid \mathcal{H}_1), \\
D^{(s)} &= t_{\mathrm{det}}^{(s)}-t_0,
\end{align}
where $P_D^{(s)}$ is detection probability, $P_{FA}^{(s)}$ is false-alarm probability, $C_{H_0}^{(s)}$ is nuisance traffic over a fixed observation horizon under no event, $C_{H_1}^{(s)}$ is alarm traffic accumulated until gateway detection under an actual event, and $D^{(s)}$ is end-to-end detection delay. The distinction between $C_{H_0}^{(s)}$ and $C_{H_1}^{(s)}$ is important because communication efficiency under nuisance conditions and communication efficiency during event-driven detection need not align.

The working hypothesis of this paper is deliberately modest: EIR can be beneficial only in a bounded operating regime characterized by (i) sufficiently separable multi-marker signatures, (ii) temporally correlated biochemical evidence, (iii) non-negligible false-alarm pressure under simpler schemes, and (iv) local inference delay $T_i$ that is not excessively large. Conversely, if the anomaly signature is weak, if the local state dynamics chatter, or if local molecular reactions are too slow, embedded inference is expected to lose its advantage despite improved selectivity.

\section{Evaluation}

\subsection{Setup}
For this study, we developed a custom stochastic system-level simulator because the question of this paper is not sequence-level DNA reaction design, but the communication-system effect of functionally abstracted local molecular inference under event-driven reporting, temporal correlation, diffusive transport, and gateway evidence integration. Existing detailed chemical or transport simulators on the one hand and detailed network simulators on the other would not directly provide this coupled abstraction level and would still require substantial custom model construction.

We instantiate the model in a simple one-dimensional vessel segment of length $L=1000\,\mu$m with gateway position $x_G=0$ and abnormality source at $x_A=750\,\mu$m. Unless otherwise stated, the simulation uses $N=50$ nanonodes, time step $\Delta t=1$~s, horizon $T_{\max}=300$~s, and abnormality onset $t_0=30$~s. Two markers are considered. Under $\mathcal{H}_0$, their mean concentrations are $(\mu_1^{(0)},\mu_2^{(0)})=(0.20,0.10)$ with noise standard deviations $(\sigma_1,\sigma_2)=(0.16,0.05)$. Under $\mathcal{H}_1$, the baseline marker amplitudes are $a_1=0.30$ and $a_2=0.12$ with spatial decay lengths $\lambda_1=250\,\mu$m and $\lambda_2=200\,\mu$m.

For EIR, we use $(w_1,w_2)=(1.0,0.90)$, so that both markers contribute strongly to the local decision while marker~1 remains slightly dominant. To make the decision more specific, marker~2 must additionally pass a gate calibrated at the 85th percentile under $\mathcal{H}_0$. Small hysteresis margins in the score ($0.03$) and in the marker-2 gate ($0.01$) make the local decision more stable and reduce rapid switching around the threshold. All strategies use edge-triggered alarming with a refractory period of 10~s, which prevents repeated emissions caused by short fluctuations. Alarm transport is modeled as one-dimensional diffusion with $D=3000\,\mu\mathrm{m}^2/\mathrm{s}$, receiver radius $5\,\mu$m, no drift, and no decay, and the gateway combines arriving alarms over time using leaky evidence integration with time constant $\tau_g^{(\mathrm{gw})}=20$~s. Finally, local thresholds are set so that only about 2\% of local states become positive under $\mathcal{H}_0$, and gateway thresholds are recalibrated for each strategy and sweep point to maintain a false-alarm level of roughly 5\%.

A key modeling assumption is that local marker concentrations evolve as variance-preserving temporally correlated processes with correlation factor $\alpha=0.85$. Under $\mathcal{H}_0$, communication cost is measured over a fixed horizon; under $\mathcal{H}_1$, it is accumulated until gateway detection. For Fig.~\ref{fig:pd_anomaly}, we sweep $(a_1,a_2)\in\{(0.03,0.012),(0.06,0.024),\allowbreak (0.10,0.040),(0.15,0.060),(0.20,0.080)\}$ to resolve the weak-event regime before saturation.

\subsection{Local state dynamics}

Before evaluating end-to-end behavior, we inspected local state dynamics because communication cost depends on how often nodes switch from OFF to ON. Under baseline $\mathcal{H}_1$, EIR has the highest mean ON fraction (0.337 versus 0.158 for RR and 0.117 for TR), so any benefit cannot be attributed to lower activity. Instead, EIR produces far fewer rising transitions per node (6.15 versus 13.90 for RR and 10.83 for TR) while keeping nodes active much longer once triggered (16.55~s versus 3.42~s and 3.25~s). Under $\mathcal{H}_0$, EIR also remains active longer (mean ON fraction 0.051 versus 0.021 and 0.022), indicating stabilized but longer-lived positive states rather than uniformly lower activity.

\subsection{System-level results}

\begin{figure}[t]
    \centering
    \begin{subfigure}{0.48\linewidth}
        \includegraphics[width=\linewidth]{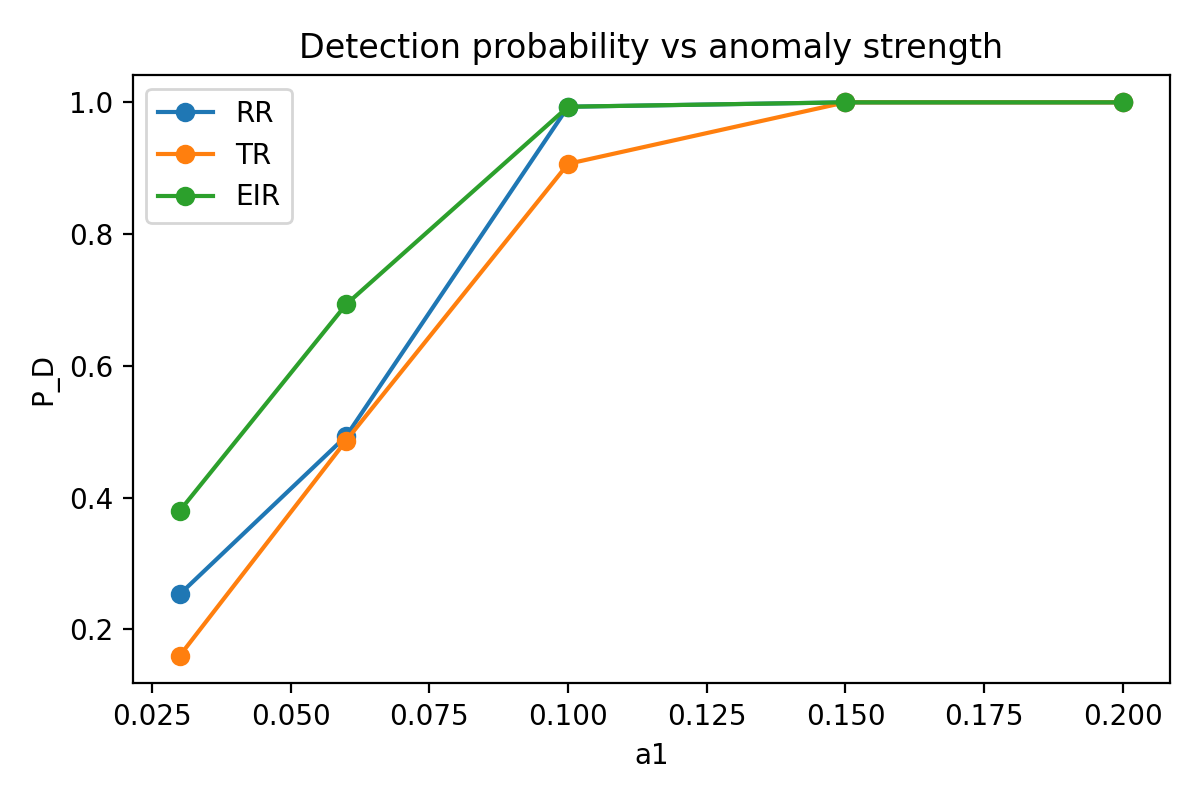}
        \caption{Detection probability versus coupled anomaly strength $(a_1,a_2)$.}
        \label{fig:pd_anomaly}
    \end{subfigure}\hfill
    \begin{subfigure}{0.48\linewidth}
        \includegraphics[width=\linewidth]{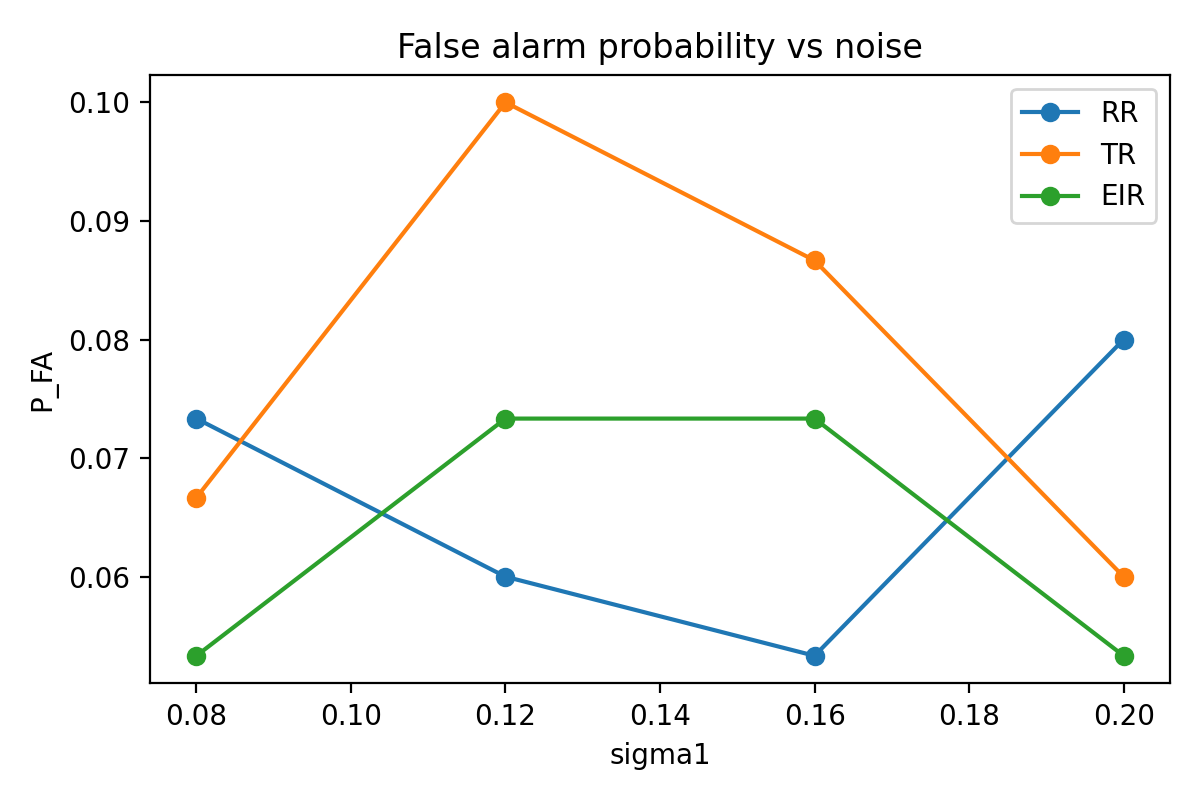}
        \caption{False-alarm probability versus noise level.}
        \label{fig:pfa_noise}
    \end{subfigure}
    \caption{Accuracy-related trade-offs of RR, TR, and EIR.}
    \label{fig:pd_pfa}
\end{figure}

Fig.~\ref{fig:pd_pfa} summarizes the accuracy-related trade-offs. The anomaly sweep in Fig.~\ref{fig:pd_anomaly} is focused on weak-to-moderate events, because stronger anomalies quickly drive all strategies into the trivial $P_D\approx1$ region. In this resolved regime, EIR shows the clearest detection advantage: at $a_1=0.03$ (with $a_2=0.012$), it reaches $P_D\approx0.38$, compared with $\approx0.25$ for RR and $\approx0.16$ for TR; at $a_1=0.06$, it rises to $P_D\approx0.69$ versus $\approx0.49$ for both RR and TR; and at $a_1=0.10$, it already reaches $P_D\approx0.99$, matching RR and exceeding TR ($\approx0.91$). Beyond $a_1\ge 0.15$, all strategies saturate at $P_D\approx1$.Thus, EIR helps mainly when anomalies are still hard to detect; once the event becomes strong enough that all strategies detect it reliably, the difference largely disappears.

Fig.~\ref{fig:pfa_noise} is again an operating-point-controlled comparison, because gateway thresholds were recalibrated for each strategy and sweep point. Within that setting, EIR remains in the same low-to-mid false-alarm band as RR and TR: for example, $P_{FA}\approx0.053$ at $\sigma_1=0.08$, $\approx0.073$ at $\sigma_1=0.12$ and $0.16$, and $\approx0.053$ at $\sigma_1=0.20$. Thus, EIR improves detection without causing an obvious false-alarm collapse, but it does not provide a universal false-alarm advantage either.

\begin{figure}[t]
    \centering
    \begin{subfigure}{0.48\linewidth}
        \includegraphics[width=\linewidth]{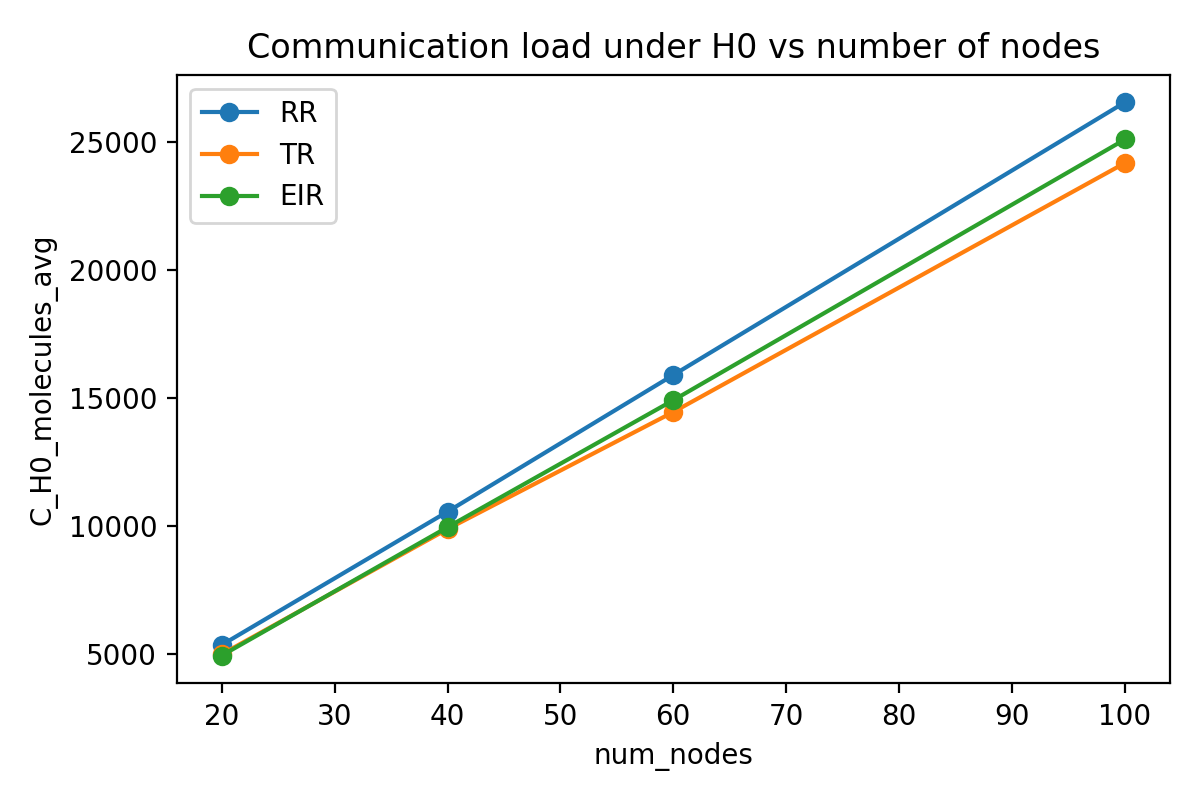}
        \caption{Nuisance traffic under H0.}
        \label{fig:comm_h0}
    \end{subfigure}\hfill
    \begin{subfigure}{0.48\linewidth}
        \includegraphics[width=\linewidth]{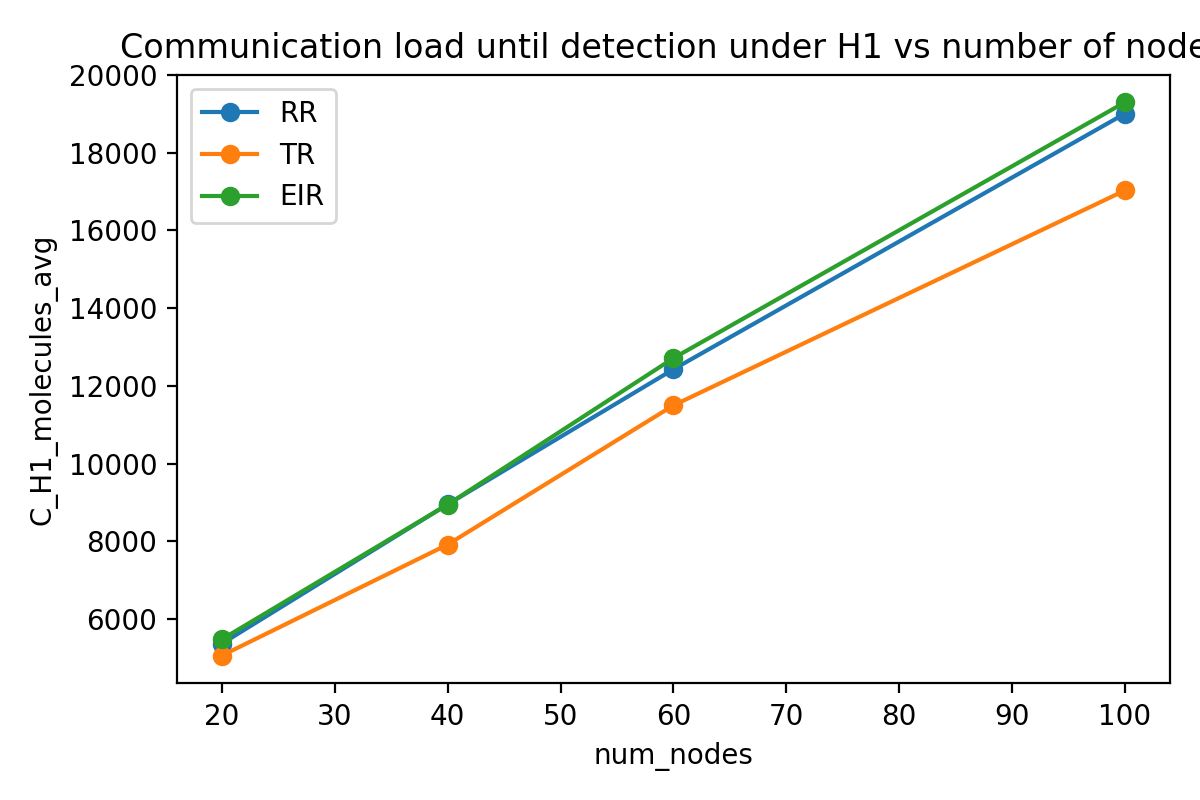}
        \caption{Event-driven traffic until detection under H1.}
        \label{fig:comm_h1}
    \end{subfigure}
    \caption{Communication-cost behavior as the network size increases.}
    \label{fig:comm}
\end{figure}

Fig.~\ref{fig:comm} shows a more nuanced communication picture. Under $\mathcal{H}_0$, nuisance traffic grows roughly linearly with network size, with RR as the most expensive strategy, TR as the cheapest, and EIR typically in between. Under $\mathcal{H}_1$, TR also remains the most communication-efficient strategy as $N$ grows: at $N=20$, the event-driven traffic until detection is about 5481 molecules for EIR, compared with 5369 for RR and 5063 for TR; at $N=100$, the corresponding values are about 19297, 19005, and 17029. EIR therefore does not provide universal traffic reduction, but often improves detection relative to RR while remaining in a comparable communication range.

\begin{figure}[t]
    \centering
    \begin{subfigure}{0.48\linewidth}
        \includegraphics[width=\linewidth]{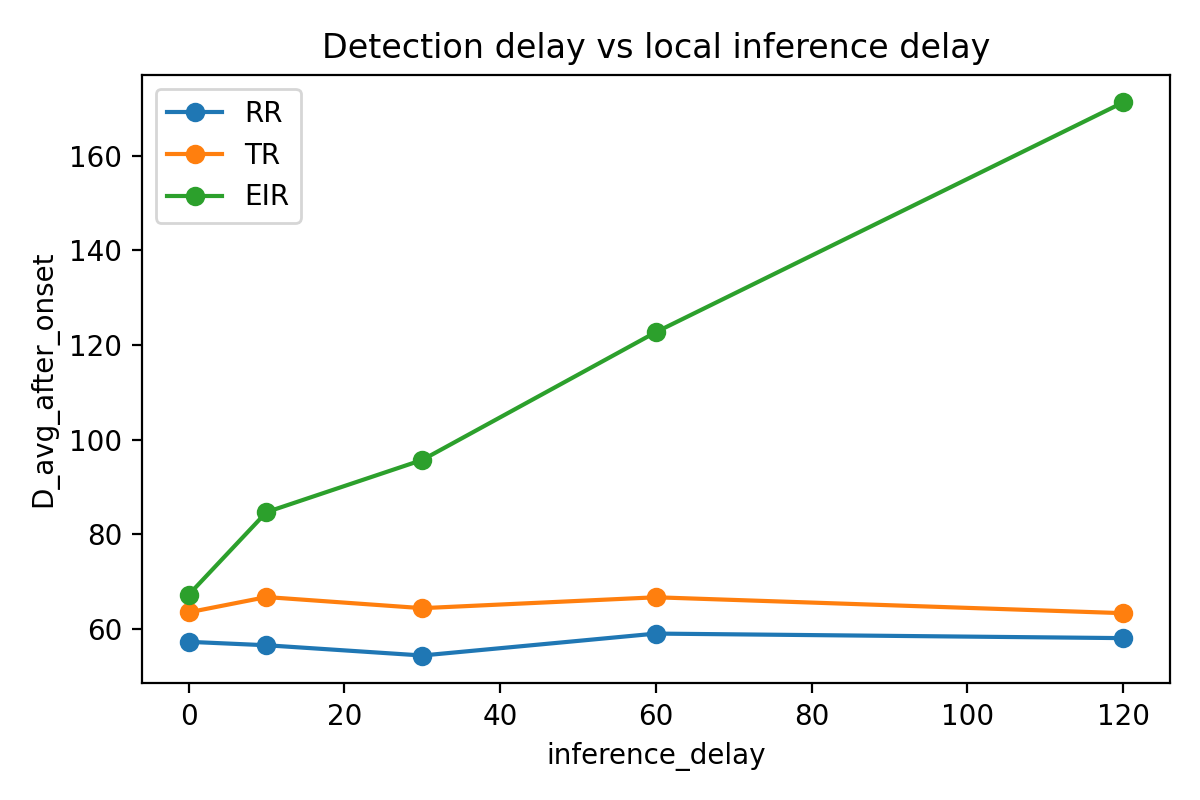}
        \caption{Detection delay versus local inference delay.}
        \label{fig:delay_ti}
    \end{subfigure}\hfill
    \begin{subfigure}{0.48\linewidth}
        \includegraphics[width=\linewidth]{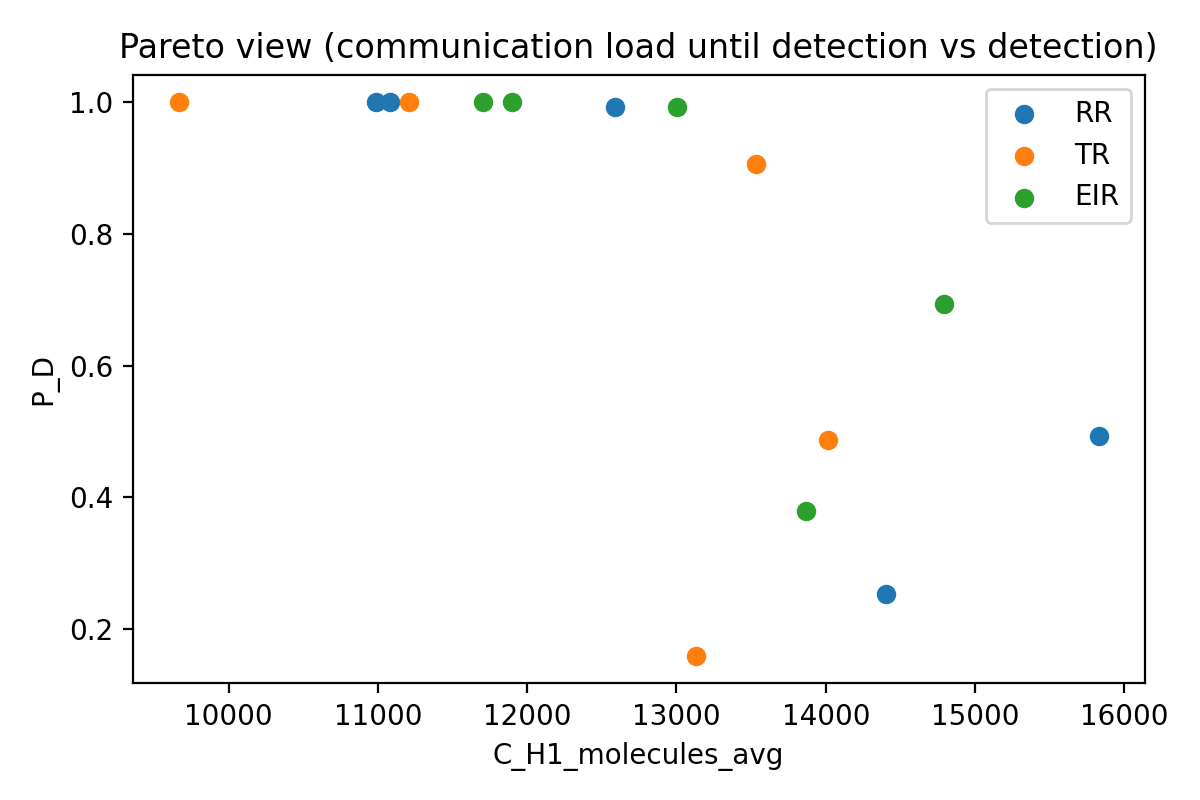}
        \caption{Pareto view of event-driven communication cost and detection probability.}
        \label{fig:pareto}
    \end{subfigure}
    \caption{Delay and Pareto-style system trade-offs.}
    \label{fig:delay_pareto}
\end{figure}

Fig.~\ref{fig:delay_pareto} shows the price of local inference. RR and TR are essentially insensitive to $T_i$, whereas the EIR delay rises from about 67~s at $T_i=0$ to about 171~s at $T_i=120$, and its communication cost increases from about 9111 to about 19093 molecules. Fig.~\ref{fig:pareto} shows the resulting bounded trade-off: at $a_1=0.06$, EIR reaches $P_D\approx0.69$ with $C_{H_1}\approx14791$, compared with $\approx0.49/15827$ for RR and $\approx0.49/14015$ for TR; at $a_1=0.10$, EIR reaches $P_D\approx0.99$ with $C_{H_1}\approx13002$, essentially matching RR in detection while staying below RR in communication cost. Once anomalies become strong, however, this advantage disappears because all strategies approach $P_D\approx1$ and TR often retains lower traffic.

Taken together, EIR is most useful for weak-to-moderate events. In that range, it can improve detection relative to RR without a large communication penalty. This gain is not free, however: EIR adds local delay, and TR often remains the cheaper strategy when abnormalities are present.

\section{Conclusion and Outlook}

We studied embedded DNA-inspired inference for multi-marker in-body nanonetworks and compared raw reporting, single-marker thresholding, and embedded inference reporting under a common system model with diffusion-based alarm transport, leaky gateway evidence integration, and variance-preserving temporally correlated marker dynamics. The simulations show that EIR is useful only in a bounded operating regime, mainly for weak-to-moderate anomalies where stabilizing local alarm dynamics improves detection. This benefit comes at the cost of additional local delay and does not make EIR uniformly preferable to TR or RR. EIR generates fewer rising alarm transitions and thus stabilizes local event signaling, while TR often remains cheaper under $H_1$. Embedded molecular inference should therefore be viewed as a communication-aware trade-off rather than a universal winner.

In future work, we will investigate richer temporal molecular integrators and distributed multi-node inference architectures in which sensing, fusion, and alarm generation are separated across interacting nanobots.

\section*{Use of AI, and Data and Software Availability}

LLMs (ChatGPT and Gemini) were used during technical development of the simulator and linguistic refinement of the overall paper; Fig.~\ref{fig:system_model} was generated with the help of Gemini NanoBanana. The author reviewed and edited all generated content and takes full responsibility for the final manuscript. Code, data, and plots are available on Zenodo (\url{https://doi.org/10.5281/zenodo.19416612}) and on GitHub (\url{https://github.com/fischesn/molecularNNsimulator}).

\printbibliography

\end{document}